\documentclass[12pt]{article}

\renewcommand\({\left(}
\renewcommand\){\right)}
\renewcommand\[{\left[}

\newcommand\eq[1]{Eq.~(\ref{#1})}

\newcommand\ee{\end{equation}}
\newcommand\be{\begin{equation}}
\newcommand\eea{\end{eqnarray}}
\newcommand\bea{\begin{eqnarray}}

\newcommand\TeV{\,\mbox{TeV}}
\newcommand\GeV{\,\mbox{GeV}}

\newcommand\mpl{M_{\rm P}}

\newcommand{\lsim}{\mbox{\raisebox{-.9ex}{~$\stackrel{\mbox{$<$}}{\sim}$~}}}

\newcommand{\gsim}{\mbox{\raisebox{-.9ex}{~$\stackrel{\mbox{$>$}}{\sim}$~}}}

\newcommand{\lesssim}{\lsim}

\def\dslash{\not{\hbox{\kern-2pt $\partial$}}}
\def\Dslash{\not{\hbox{\kern-4pt $D$}}}
\def\Oslash{\not{\hbox{\kern-4pt $O$}}}
\def\Qslash{\not{\hbox{\kern-4pt $Q$}}}
\def\pslash{\not{\hbox{\kern-2.3pt $p$}}}
\def\kslash{\not{\hbox{\kern-2.3pt $k$}}}
\def\qslash{\not{\hbox{\kern-2.3pt $q$}}}

\newcommand\sub[1]{_{\rm #1}}

\newcommand\meff{m\sub{eff}}
\newcommand\meffs{m\sub{eff}^2}

\begin{document}

\begin{titlepage}

\title{Curvaton and QCD Axion in Supersymmetric Theories}

\author{Eung Jin Chun$^1$, Konstantinos Dimopoulos$^{2,3}$
and David H. Lyth$^3$ \\[1ex]
$^1${\small\it Korea Institute for Advanced Study}\\
{\small\it 207-43 Cheongryangri-dong, Dongdaemun-Gu}\\
{\small\it Seoul 130-722, Korea}\\
$^2${\small\it Institute of Nuclear Physics, National Center for Scientific
Research: `Demokritos',}\\
{\small\it Agia Paraskevi Attikis, Athens 153 10, Greece}\\
$^3${\small\it Physics Department, Lancaster University,
Lancaster LA1 4YB, U.K.}
}

\date{April, 2004}

\maketitle

\abstract{ \normalsize A pseudo Nambu-Goldstone boson as curvaton
avoids the $\eta$-problem of inflation which plagues most curvaton
candidates.
We point out that a concrete realization of the curvaton mechanism with a
pseudo Nambu-Goldstone boson can be found in the supersymmetric
Peccei-Quinn mechanism resolving the strong CP problem. In the
flaton models of Peccei-Quinn symmetry breaking, the angular
degree of freedom associated with the QCD axion can naturally be a
flat direction during inflation and provides successful curvature
perturbations. In this scheme, the preferred values of the axion
scale and the Hubble parameter during inflation turn out to be
about $10^{10}$ GeV and $10^{12}$ GeV, respectively.
Moreover, it is found that a significant isocurvature component,
(anti)correlated to the overall curvature perturbation, can be generated,
which is a smoking-gun for the curvaton scenario.
Finally, non-Gaussianity in the perturbation spectrum at
potentially observable level is also possible. }

\thispagestyle{empty}

\end{titlepage}

\section{Introduction}

The primordial curvature perturbation is caused presumably by some
scalar field, which acquires its perturbation during inflation.
For a long time it was generally agreed that the
curvature-generating field would be the inflaton \cite{treview}.
Then it was suggested  instead  that this field is some other
`curvaton' field \cite{LW,moroi} (see also \cite{also}), which
generates the curvature perturbation only when its  density
becomes a significant fraction of the total. The curvaton proposal
has received enormous attention because it opens up new
possibilities both for model-building and for
observation.\footnote {More recently still it has been suggested
that the curvaton acts by causing inhomogeneous reheating
\cite{dzlev} or through a preheating mechanism \cite{steve}.}

In all cases,  the field responsible for the curvature perturbation
must be  light during
inflation, in the sense that its effective mass $\meff$ is much less than
the Hubble parameter $H$. (To be precise, we need during inflation
$\meffs \lsim 0.1 H^2$,
so that the spectral tilt $1-n\simeq \frac23 \meffs/H^2$ satisfies the
observational constraint.)
If the responsible field is a curvaton, it
 should preferably also remain light
after inflation, until $H$ falls below the true mass \cite{CD}.
These requirements are
 in mild conflict with the generic expectation from supergravity,
that the mass-squared of each scalar field  will be at least of order $H^2$
\cite{DRT}. This is the famous $\eta$-problem of inflation.
Ways have been proposed to keep
the responsible  field sufficiently light \cite{treview},
the most straightforward
of them being to make it a pseudo Nambu-Goldstone boson (PNGB)
(for the curvaton see \cite{LW,pngb}).

For economy, and also to facilitate contact with observation, one would
like a candidate for the responsible field to be one that is present in
an already-existing model, designed for some purpose other than
the generation of  the curvature perturbation. Several such candidates have
been proposed \cite{pngb,andmore,pqcurv,kkt,McD}, but in general they do not
come with a satisfactory mechanism for keeping
the curvaton sufficiently light.
{\em The purpose of this paper is to suggest a natural
curvaton candidate, which is present in flaton models of
Peccei-Quinn symmetry breaking, and which is a PNGB.}

The Peccei-Quinn (PQ) symmetry  provides a nice solution to the
strong CP problem \cite{pq}. It  is  an anomalous global symmetry,
$U(1)_{PQ}$, spontaneously broken at an intermediate scale
$f_{PQ}$ with preferred value around $10^{12}$ GeV providing
enough dark matter (axion) of the universe. Its PNGB is the axion,
which must be extremely light to satisfy the observational
constraints \cite{INV}. In the context of supersymmetry (SUSY),
two complex fields (or more) are needed to implement a global
symmetry due to holomorphicity of the superpotential (unless the
symmetry in question is
R--symmetry). Hence, in SUSY, to spontaneously break the  PQ
symmetry one needs at least two complex fields, corresponding to
two radial
 and two real angular degrees of freedom. One of the latter is
 the axion, and the other is our curvaton candidate.\footnote
{By `degree of freedom' we mean as usual a
normal mode of the coupled oscillations of the fields,
which after quantization corresponds to a particle species.}
We can define a symmetry (explicitly broken of course) acting
only on the combination of phases which corresponds to the curvaton,
 and then the curvaton is the  PNGB of that symmetry.

To understand how our model works, recall first that
two  fundamentally different paradigms exist for the PQ symmetry
breaking fields.
In one of them,  the potentials involve
only renormalizable terms in the superpotential plus small soft
SUSY breaking terms, leading to a normal Mexican-hat potential whose
height and width have the same  scale $f_{PQ}$.
According to this paradigm,
 PQ symmetry breaking persists in the limit of unbroken SUSY
so that the axion has a well-defined supersymmetric scalar partner
called the saxion (as well as a fermionic partner called the axino).
The saxion is the only light degree of freedom
 apart from the axion, its mass
coming from SUSY breaking and being of order $\TeV$ for
gravity-mediated SUSY breaking. The other two degrees of freedom
 are heavy with mass of order $f_{PQ}$.

According to the other  paradigm, the PQ symmetry breaking fields
are instead flaton fields \cite{flaton}, so-called because they
are symmetry-breaking fields, which correspond to flat directions
of the potential (i.e., directions in which the quartic term is
negligible). Their  potential contains only soft SUSY breaking
terms and non-renormalizable terms, which means that it is very
flat. In the limit of unbroken SUSY there would be no spontaneous
breaking of the PQ symmetry. A nice feature of this paradigm is
that the intermediate axion scale $f_{PQ}$ is not put by hand but
is generated through a geometric mean of the SUSY breaking scale
and the Planck scale.
 The other degrees of freedom
accompanying the axion  get their mass only from SUSY breaking,
making them of
 order $\TeV$ for gravity-mediated SUSY breaking.
In this paper, we point out that this kind of model
contains a natural curvaton candidate as the angular degree of
freedom. The radial degrees of freedom are less suitable for this
purpose, because in the early Universe they presumably acquire the
mass of order $H$ which is generic \cite{DRT} for scalar fields in
a supergravity theory. In contrast, a mass of order $H$ is
unlikely to be generated for the angular degree of freedom,
because it would correspond to a generalized $A$-term which is
forbidden if the fields responsible for the energy density are
charged under certain symmetries \cite{DRT}.
We will
analyze how the model parameters like the axion scale, the Hubble
parameter and the curvaton decay temperature
are constrained to produce successful curvature perturbations.

\section{Supersymmetric realization of the PQ and curvaton mechanisms}

The minimal model of the type that we are considering is a
variation of the non-renormalizable superpotential
\cite{MSY,choi,laza}, which introduces two fields $P$ and $Q$
charged under PQ symmetry and one singlet $S$ allowing the
following superpotential
\begin{equation} \label{WPQ}
 W = h {P^{n+1} \over M_P^n} H_1 H_2 +  \lambda_1 {P Q S^{n+1} \over
 M_P^n} + \lambda_2 {S^{n+3}\over M_P^n} \,
\end{equation}
where the $U(1)_{PQ}$ and $Z_N$ charges are assigned as
\begin{equation}
\begin{array}{rccccc}
  & H_1 & H_2 & P & Q & S\vspace{.3cm}\cr
 U(1)_{PQ}:\quad &  {1\over2} (n+1)  & {1\over2} (n+1)   & -1 & +1 & 0 \cr
 Z_N: \quad& 1 & 1 & 1 & \alpha^2 & \alpha   \cr
\end{array}
\end{equation}
with $\alpha\equiv e^{2\pi i/N}$ and $N=n+3$. The field $S$  is
supposed to have a {\it negative} soft mass-squared  of order the
gravitino mass $m_{3/2}$. This  forces all the fields $P$, $Q$ and
$S$ to develop  the vacuum expectation values of the order
\begin{equation}
 v
 \sim \left( m_{3/2} M_P^n \right)^{1/n+1} \,.
\label{vacv}
\end{equation}
This sets the intermediate PQ scale
\begin{equation}
f_{PQ} \sim 3\times 10^{10},\;\; 1\times
10^{13},\;\;3\times10^{14}\;\mbox{ GeV}
\end{equation}
for $n=1,2,3$, respectively, and
$m_{3/2} = 300$ GeV. The negative mass-squared of the $S$ field
can come from the renormalization effect \cite{MSY} or from the
initial condition of soft masses.  We do not assume the negative
mass-squared for the $P$ and $Q$ fields, which  would drive the
scalar potential unbounded from below when some of the fields $P$,
$Q$ or $S$ are set to zero.

Note that the superpotential of Eq.~(\ref{WPQ}) provides a natural
solution to the $\mu$ problem \cite{CKN} as the first
non-renormalzable term leads to the right order of magnitude for
the Higgs mass parameter;
$$ \mu =h  {v_P^{n+1} \over M_P^n} \sim m_{3/2}.  $$
The above interaction will be the main source of the decays of the
particles  corresponding to $P$, $Q$ or $S$ into the standard
(s)particles.

The model under consideration contains six gauge singlet scalar
particles. Three radial fields get masses of order $m_{3/2}$ from
soft supersymmetry breaking. Among three phase fields,
$\varphi_P$, $\varphi_Q$ and $\varphi_S$, one combination,
$\varphi_P+\varphi_Q-2\varphi_S$, gets the mass of order $m_{3/2}$
as the radial fields.  The combination, $\varphi_P-\varphi_Q$, is
the axion. The other combination is our curvaton candidate,
$\varphi_P+\varphi_Q + \varphi_S$, which becomes massive due to
the soft supersymmetry breaking $A$ term;
$$ V_{soft} = A \lambda_1 {P Q S^{n+1} \over
 M_P^n} + A \lambda_2 {S^{n+3} \over
 M_P^n} + h.c. \;,$$
inducing  also in the vacuum a curvaton  mass of order $m_{3/2}$.
(For a detailed calculation in a similar model of the mass
spectrum, see \cite{comelli}.) Note that, in our scheme, the
curvaton parameters like its amplitude and decay temperature are
fixed by the PQ scale and the $\mu$ term, as will be shown
explicitly.

\section{The potential in the early Universe}

In order to consider the basic features of the curvaton field $\sigma$
associated with the QCD axion,  let us simplify the original superpotential
term in Eq.~(\ref{WPQ}) as follows;
\begin{equation} \label{Wphi}
 W(\phi) = {\lambda \over n+3}{\phi^{n+3} \over M_P^n}\,,
\end{equation}
where  the complex field $\phi$ may be thought of as containing
the phase field $\sigma$ and one radial component. This would be
literally correct if the vacuum expectation values (VEVs) of $P$,
$Q$ and $S$ happened to be  equal, allowing the parameterization,
\begin{equation}
\phi\equiv |\phi|e^{i \theta}=|\phi|\exp(i\sigma /\sqrt{2} v)
\label{theta}
\end{equation}
with $v=\langle \phi \rangle$ and \mbox{$\sigma\equiv\sqrt 2\, v \theta$}.
(As  noted in the  Conclusion,
our curvaton model can also be implemented using only $\phi$, though
one then loses the connection with the QCD axion.)

With the above superpotential (\ref{Wphi}), the general scalar
potential in the early universe can be written as
\begin{equation}
 V = (3C_\phi H^2 - m_\phi^2) |\phi|^2 + \left[(C_A H +
 A){\lambda\over n+3}
 {\phi^{n+3} \over M_P^n} +h.c.\right]+ \lambda^2 { |\phi|^{2n+4} \over
 M_P^{2n} }\,,
\end{equation}
where $m_\phi$ and $A$ are soft supersymmetry breaking masses at
zero temperature. Note that we have put negative mass-squared for the
$\phi$ field at zero temperature to generate the PQ scale as discussed
previously.

The terms with $H$ are expected to arise from the supersymmetry
breaking effect during and after inflation \cite{DRT}, except
during  radiation domination \cite{p03moroi}. The generic
expectation for $C_\phi$ is $|C_\phi|\sim 1$, and we are going to
assume $C_\phi \sim -1$ so that there is PQ symmetry breaking. As
is well known, a value
 $|C_A|\ll 1$ can be achieved provided that relevant field values
are small on the Planck scale, if a symmetry forbids low-order
 terms in the K\"{a}hler potential which are linear in the field responsible for
the energy density \cite{DRT}.
 To be safe though, we need
an estimate of the actual value of $C_A$,
considering for completeness also $C_\phi$.

During inflation, the
supergravity potential is taken to be generated by a single $F$--term
$F_I$;
\begin{equation}
V= e^{K/\mpl^2} \(|F_I |^2 - 3{|W|^2 \over M_P^2} \) = 3 H^2 M_P^2\;,
\end{equation}
where
\be
|F_I|^2  \equiv \( \partial_I W+ W\partial_I K/M_P^2 \) K^{I I^*}
\( \partial_I W+ W\partial_I K/M_P^2 \)^*
\ee
and $K^{II^*}$ is an element of the inverse of the  K\"{a}hler metric
$K_{IJ*}\equiv \partial_I \partial_{J*} K$ with $J$ running over
all of the scalar fields.
Depending on the model, the  field $I$ might be the inflaton, the waterfall
field or some other field \cite{treview}.

We make the following
{\it assumptions};
(i) in the K\"{a}hler potential,
terms linear in $I$ are negligible (this
 can be a consequence of a  global or gauge symmetry),
(ii) the field $I$  is much smaller than the Planck mass; $I
 \ll M_P$, as is true in most inflation models,
and (iii) the field $\phi$ is also much smaller than $M_P$; $\phi
\ll M_P$. The last two conditions allow us to make a perturbative
expansion of the potential in terms of $I$ and $\phi$.
Under these assumptions,
we write the most general K\"ahler potential and superpotential :
\begin{eqnarray}
 K &=& I^\dagger I + \phi^\dagger \phi + {c\over M_P^2} I^\dagger I
 \phi^\dagger \phi + \left[ {d\over M_P^3} I^\dagger I W(\phi) + h.c.
 \right]
 \nonumber\\
 W &=& W(I) + W(\phi)
\end{eqnarray}
neglecting terms with higher powers of $I^\dagger I$,
$\phi^\dagger \phi$ and $W(\phi)$ which give more suppressed
contributions.
  The conditions (ii) and (iii) imply that $F_I \approx \partial_I W \approx
\sqrt{3} H M_P$ and $W/M_P \equiv  \alpha I H \ll F_I$ with
$\alpha \lesssim 1$. Then, we have
\begin{equation}\label{Cs}
 C_\phi \approx (1-c),\quad
 C_A \approx \sqrt{3}\,\Big[1-c(n+3)-\sqrt{3}\alpha\Big] {I\over M_P} - 3 d{H\over M_P}\,.
\end{equation}
As advertised, one obtains $|C_\phi|\sim 1$ but $|C_A|\ll 1$
\cite{DRT}.

If $H$ replaces $m_\phi$, \eq{vacv} becomes $v\sim (H M_P^n)^{1/n+1}$,
which leads to the following;
\begin{equation} \label{vsare}
v \sim  \cases{
   (H M_P^n)^{1/n+1}  & \mbox{during inflation}, \cr
   [\mbox{max}(H,m_{3/2}) M_P^n]^{1/n+1} &
   \mbox{after inflation}, \cr
   (m_{3/2} M_P^n)^{1/n+1} = f_{PQ} & \mbox{after reheating}. \cr }
\end{equation}
While  the radial degree of freedom  follows such a thermal
history, the corresponding phase degree of freedom, $\sigma$, will
enjoy the following potential of the pseudo Nambu-Goldstone boson (PNGB)
type \cite{pngb}:
\begin{equation}
V(\sigma) \approx (C_A H + A ) v^3\left(\frac{v}{M_P}\right)^n\,
\left[1-\cos\left(\frac{\sigma}{v}\right)\right] .
\end{equation}
which is very flat during and after inflation.
{}From Eq.~(\ref{vsare}), one finds
\begin{equation} \label{msare}
m_\sigma^2  \sim
(C_AH+A)\max\{H, m_{3/2}\}\sim
\cases{
   C_A H^2   & \mbox{during inflation}, \cr
   \mbox{max}( C_A H^2, AH, A m_{3/2}) & \mbox{after inflation}, \cr
   A m_{3/2} & \mbox{after reheating,} \cr }
\end{equation}
where we assumed that, during inflation \mbox{$H_*>A/C_A$}.

\section{The curvaton cosmology }

It is now obvious that the field $\sigma$ will naturally realize the curvaton
paradigm.
Taking Eq.(\ref{Wphi}) literally, $\sigma$ can be parameterized
as \mbox{$\theta \equiv \sigma/\sqrt{2} v$}, that is,
\mbox{$\phi= ve^{i\theta}$} [cf. Eq.~(\ref{theta})].
To obtain simple predictions, let us
assume that the value of $\sigma$ is completely randomized during inflation,
so that in our part of the Universe $\theta$ at horizon exit
has some value $\theta_0 < 2\pi$.
 (We  will come back to this point later).

It
remains constant as far as $H \gg m_\sigma$ and the
oscillation starts at $H \sim m_{3/2}$.  Finally, the curvaton
decays and thermalizes at $H \sim \Gamma_\sigma$ which is determined to
be quite late due to its axion-like couplings.
The interaction of $\sigma$ with ordinary particles is governed by
the effective $\mu$ term in Eq.~(\ref{WPQ}) leading to
$\mu = h f_{PQ}^{n+1}/M_P^n \sim m_{3/2}$.  Thus,
the decay rate of $\sigma$ into two Higgs field is
\begin{equation} \label{Gs}
\Gamma_\sigma \approx{(n+1)^2 \over 4\pi} {m_{3/2}^3\over
f_{PQ}^2}\,,
\end{equation}
where we put $\mu=m_\sigma=m_{3/2}$.  Taking the approximation of
$\Gamma_\sigma \sim m^3_{3/2}/f_{PQ}^2$ and
$f_{PQ} \sim (m_{3/2} M_P^n)^{1/n+1}$,
we get the decay temperature of $\sigma$ as follows;
\begin{equation}
T_{\rm dec} \sim \cases{ 160 \mbox{ GeV}\quad \mbox{for $n=1$} \cr
0.6 \mbox{ GeV}\quad \mbox{for $n=2$} \cr 30 \mbox{ MeV}\quad
\mbox{for $n=3$} \cr}
\end{equation}
with $m_{3/2} =300$ GeV.  Here we have restricted ourselves to
$n\leq 3$ for which the decay occurs well before the
nucleosynthesis occurring around $T_{\rm BBN} \sim 1$ MeV.

The density perturbation driven by the curvaton is given by \cite{LW}
\begin{equation} \label{zetais}
\zeta = {1\over \pi} {r \over 4+ 3r} {H_* \over \sigma_*}
\end{equation}
where $r$ is the density ratio of $\rho_\sigma/\rho_\gamma$ at the time of
the $\sigma$ decay:
\begin{equation} \label{ris}
  r \approx {1\over 6} \left(  H_{\bar{m}} \over \Gamma_\sigma  \right)^{1/2}
         \left( \sigma_* \over M_P \right)^2  .
\end{equation}
Here $H_{\bar{m}} \equiv \mbox{min}(m_\sigma, \Gamma_I)$ \cite{CD}
where $\Gamma_I$ is the inflaton decay rate determining the reheating
temperature. Note that, in our model, the values of $r, H_*$ and $\sigma_*$
are determined by dimensionality of the superpotential in Eq.~(\ref{Wphi}),
which are related to the axion scale, $f_{PQ}$.

Inserting Eqs.~(\ref{vsare}) and (\ref{Gs}) to
Eqs.~(\ref{zetais}) and (\ref{ris}), we can determine
the inflation Hubble parameter $H_*$
and the density ratio $r$ which reproduce the observed value of
$\zeta \approx 5\times10^{-5}$.
First, the value of $r$ can be calculated
from the relation:
\begin{equation}\label{req}
 {r^{n+2 \over 2} \over 1 + {3\over4} r} \sim
 \frac{4\pi\zeta\, \theta_0^{n+1}}{6^{n/2}}
 \left( H_{\bar{m}} \over m_{3/2}\right)^{n\over4}
 \left( M_P \over m_{3/2} \right)^{n^2 \over 2(n+1)}
\end{equation}
and then it fixes the inflation Hubble parameter as
\begin{equation}\label{zeta}
 {H_* \over M_P} \sim \left(4 \pi \zeta\, \theta_0  \right)^{n+1 \over n}
  \left( {1+ {3\over4}r \over r}   \right)^{n+1 \over n}  .
\end{equation}
For a typical set of input parameters;
$m_{3/2} = 300$ GeV and  $H_{\bar{m}}=1$ GeV,
we get the following values for $H_*/M_P$ and $r$;
\begin{equation}\label{results}
 \Big({H_* \over M_P},\; r\Big) \sim
\left\{\begin{array}{lll} (8\times 10^{-7}\,\theta_0^{-2/3}\;,
&\;1.4\,\theta_0^{4/3}) &
{\rm for}\; n=1\\
(1\times 10^{-5}\,\theta_0^{3/2}\;, &\;6\times 10^5\,\theta_0^3) &
 {\rm for}\; n=2\\
(4\times 10^{-5}\,\theta_0^{4/3}\;, &\;2\times
10^8\,\theta_0^{8/3}) &
 {\rm for}\; n=3
 \end{array}\right.
\end{equation}
which is valid for the typical case of $\theta_0\lsim 1$, i.e.
$r\lesssim1$ for $n=1$ and
$r\gg1$ for $n=2,3$. The exact behaviors of $(H_*/M_P,r)$
depending on $\theta_0$ are shown in Figures 1, 2 and 3.

The value of $r$ has to be large enough to avoid excessive
non-Gaussianity in the density perturbation spectrum. The
constraint from the WMAP observations reads: $r>9\times10^{-3}$
\cite{wmap}. Enforcing this constraint in Eq.~(\ref{req}) we
obtain the bound
\begin{equation}
\theta_0>\frac{0.232}{(103.3\,\pi\zeta)^{1/(n+1)}}
 \left(H_{\bar{m}} \over m_{3/2}\right)^{-{n\over4(n+1)}}
 \left( M_P \over m_{3/2} \right)^{-{1\over 2}({n\over n+1})^2}.
\label{thetabound}
\end{equation}

Suppose first that we live in a typical part of the Universe, then
\mbox{$\theta_0\sim{\cal O}(1)$} (see appendix for a discussion on
the randomization of the curvaton field). In this case the value
of $r$ is always large enough to avoid excessive non-Gaussianity.
For $n=2,3$, $H_*$ becomes
barely compatible with the limit $H_*/M_P < 10^{-5}$ coming from the
observational bound on the primordial
tensor perturbation. (Actually, the curvaton model requires a
somewhat smaller limit \cite{p02kostas}, assuming that inflation
is of the slow-roll type.) The case with $n=1$ then gives most
satisfactory result with $H_* \sim 10^{12}$ GeV. In this case, the
curvaton decay temperature is $T_{\rm dec} \sim 100$ GeV and the
resulting entropy dumping is not significant (it is comparable to
the preexisting one), so that the usual cosmological properties
are retained. That is, the conventional dark matter candidate, the
lightest neutralino or the axion with $f_{PQ} = 3\times 10^{10}$
GeV, is equally acceptable within our curvaton context, and the
usual baryogenesis mechanism can also work.
Note, also, that in this case we can have substantial non-Gaussianity in
the density perturbation spectrum. Indeed, from Eq.~(\ref{results}) we
readily obtain \cite{iso1}
\begin{equation}
f_{\rm NL}=\frac{5}{4r}\simeq 0.9\,\theta_0^{-4/3}.
\label{fNL}
\end{equation}
The largest value of the above corresponds to the smallest possible
value of $\theta_0$, which is determined by Eq.~(\ref{zeta}),
when we demand that \mbox{$H_*/M_P<10^{-5}$}. With a little algebra it can be
shown that \mbox{$(\theta_0)_{\rm min}\simeq 0.34\sqrt{r}
$}, which
gives \mbox{$(f_{\rm NL})_{\rm max}\simeq{\cal O}(10)
$}. Such a value of $f_{\rm NL}$ may be observable in the near future.

In the above we assumed a reheating temperature \mbox{$T_{\rm
reh}\sim 10^9$GeV}, which saturates the constraint of gravitino
overproduction. This constraint is important, in the case when
$n=1$, because the entropy production by the curvaton decay is not
enough to dilute the gravitinos. However, a lower $T_{\rm reh}$
corresponds to a smaller $H_{\bar{m}}$, which results in a smaller
$r$, according to Eq.~(\ref{req}). This, in turn, results in a
larger $H_*$ [cf. Eq.~(\ref{zeta})], which again violates the
upper bound on $H_*$. On the other hand, a higher $T_{\rm reh}$
corresponds to the range \mbox{1 GeV $<H_{\bar{m}}\leq
m_{3/2}$}, which, for a given $n$, can increase $r$ somewhat.
If \mbox{$r\gg 1$} then the entropy production by the decay of the
curvaton relaxes the gravitino constraint. However, as shown by
Eq.~(\ref{zeta}), $H_*$ loses its sensitivity to a large $r$, even
though, as can be seen in Figure~1, $H_*$ may be reduced down to
$10^{-7}M_P$ for \mbox{$H_{\bar{m}}\sim m_{3/2}$}.\footnote{Note,
here, that there is also a lower
bound on $H_*$, as was derived in \cite{DLbound}. In our case it
can be shown that, because $v$ during inflation is different that
in the vacuum, this bound reads
\mbox{$H_*>(v_*/f_{PQ})^{4/5}10^{-12}M_P$} \cite{yeinzon}.
Using, Eqs.~(\ref{vacv}) and (\ref{vsare}) this bound can be recast as
\mbox{$H_*>[10^{-15(n+1)}(M_P/m_{3/2})]^{\frac{4}{5n+1}}M_P$}.
The tightest case corresponds to \mbox{$n=1$}, which gives
\mbox{$H_*>10^{-9}M_P$}. Clearly, our results in Eq.~(\ref{results})
satisfy well this bound for \mbox{$\theta_0\sim{\cal O}(1)$}.}

Alternatively, we may allow the possibility that
$\theta_0$ is untypically small in our region. Then the value of
$H_*$ needed to generate the observed density perturbation is
reduced by a factor $\theta_0^\frac{n+1}{n}$ for $n=2,3$.
Now the cases \mbox{$n=2,3$} become viable, but the $n=1$ is not
because it violates the  $H_*\lsim 10^{-5}\mpl$ bound
[cf. Eq.~(\ref{results})].
Moroever, for $n=2$ and 3 we now have the fascinating possibility
that the curvature perturbation has the observed magnitude only
because we live in a special part of the Universe. (To be precise,
the bound $H_*\lsim 10^{-5}\mpl$ makes this state of affairs
inevitable for $n=2,3$)
This is actually a quite common outcome of curvaton models
\cite{pngb}, and provides a good example of how a theoretical
model can make anthropic considerations mandatory (in absence of a
model to fix $\theta$ dynamically to a small value).

In this model, Peccei-Quinn symmetry is broken during inflation
and is not restored. No axionic strings are formed, and axions are
produced by the oscillation of the homogeneous axion field. Their
density is \cite{INV}
\be
\Omega\sub a = (1\ {\rm to}\ 10) \( \frac{f_{PQ}
}{10^{12\GeV}} \)^{1.2}
\( \frac{N\theta\sub a}{\pi} \)^2 D
,
\ee
where $N$ is the multiplicity of the axion vacuum and
$D$ is the dilution coming from entropy production after axion
creation. (In our model, $D=1$ for the most interesting case $n=1$.)
The uncertainties are big, but it is clear that the
axions can be the dark matter ($\Omega\sub a = 0.25$) for
$\theta\sub a$ roughly of order 1.

The perturbation $\delta\theta\sub a$ will generate a matter isocurvature
perturbation $S\sub m$, which should be $\lsim 0.1\zeta$ to be compatible with
observation. It is given by
\bea
S\sub m &=& \frac 13 \Omega\sub a \frac {\delta \rho\sub a}{\rho\sub a} \\
&=&  \frac23  \Omega\sub a \frac {\delta \theta\sub a}{\theta\sub a}
.
\eea
But $\delta \theta\sub a\sim \delta\theta_0$ because both are
generated from the vacuum fluctuation.
Therefore,
in view of Eq.~(\ref{zetais}), we find

\be
\frac{S\sub m}{\zeta}  \sim
 (1\ {\rm to}\ 10) \( \frac{f_{PQ}
}{10^{12}\GeV} \)^{1.2}
\( \frac{N}{\pi} \)^2 \theta\sub a \theta_0
\left(\frac{4+3r}{3r}\right)
 D
, \ee which can be $\lsim 0.1$ as required by present observation.
On the other hand it could be observable in the future.

\section{Discussion and Conclusions}

Within the curvaton paradigm, we
 have proposed a new candidate for the field which causes the
curvature perturbation. We believe that it is one of the most
attractive  candidates yet   proposed for that field. Being a
field that is part of the flaton realization of PQ
symmetry-breaking, and it can easily be kept light during and
after inflation. In a typical part of the universe, the observed
magnitude $\sim 10^{-5}$ of the curvature perturbation comes  from
a  modest and quite reasonable hierarchy between the inflationary
Hubble parameter and the vacuum expectation value  of the
Peccei-Quinn field. The predicted values of the axion scale and
the Hubble parameters are $10^{10}$ and $10^{12}$ GeV,
respectively, corresponding to $n=1$.  In this case, the entropy
dumping due to the curvaton decay is negligible so that the
conventional cosmological predictions concerning  dark matter
components and  baryogenesis remain unchanged.  The model can
generate two kinds of isocurvature pertubations. One is an axion
isocurvature perturbation, uncorrelated with the curvature
perturbation.  The other is an isocurvature perturbation in some
other kind of dark matter, or in the baryonic matter, which is
produced either  before or at curvaton decay. As described in
\cite{iso1,iso2}, such a perturbation is generic to curvaton
models, and is fully correlated or anti-correlated with the
curvature perturbation.\footnote{Correlation or anti-correlation of the CDM
\{baryon\} isocurvature perturbation depends on whether CDM
creation  \{baryogenesis\} takes place (just) before curvaton
decay or due to the curvaton decay itself. Note, that there is no
residual isocurvature perturbation if CDM creation
\{baryogenesis\} occurs after curvaton decay.}
The detection of any such isocurvature
perturbation would be evidence in support of the curvaton model.

Finally, we have shown that, in this case, non-Gaussianity at at
an observable level (with \mbox{$f_{\rm NL}\sim {\cal O}(10)$}) is
possible.    Our  curvaton candidate is  a PNGB, corresponding to
an angular degree of freedom associated with the QCD axion. As a
consequence, it can avoid the usual mass of order $H$ during (and
after) inflation, because supersymmetry breaking affects the mass
only through $A$ terms, which can be controlled  by appropriate
symmetries.

Our model of PQ symmetry breaking is similar to  one already
extensively investigated \cite{comelli,hangbae}, except that we
take the PQ symmetry to be spontaneously broken throughout the
history of the Universe. In contrast, the investigations of
\cite{comelli,hangbae} assume that the symmetry is initially
unbroken, leading to thermal inflation. These possibilities
correspond respectively to radial masses-squared of order $\mp
H^2$, and the two signs should be deemed equally likely in the
absence so far of a string-theoretic prediction. The unbroken
paradigm has a very different cosmology from the one that we are
adopting, producing in particular copious saxion- or axino-like
particles which may  decay into relativistic axions whose energy
density is enough to affect nucleosynthesis \cite{comelli}.
Also, the lightest axino-like particle may be the lightest supersymmetric
particle \cite{hangbae}, providing a more natural implementation of the Cold
Dark Matter scenario that was originally  proposed \cite{leszek} in the
context of non-flaton models. None of this occurs within our paradigm.

As it invokes the axion, our model, as it stands, is open to the
criticism that the axion mass is implausibly small. Indeed,
in the kind of models that we have considered where the axion is
an angular part of a complex field, a non-renormalizable
 term  in the potential  with dimension $d$
generically breaks PQ symmetry and  contributes
 to the axion mass an amount of order $\sim v^{(d-2)/2}$ in
Planck units. To keep the axion mass to the required value
of order $10^{-30}\mpl$, terms up to $d\sim 12$ must  respect
the PQ symmetry. (We take $v\sim f_{PQ} = 10^{12}\GeV$ for an estimate.)
A widely-discussed possibility to ensure this is to impose
a $Z_n$ subgroup of the PQ symmetry \cite{lukas}.\footnote
{Imposing the full PQ symmetry of course works, but exact continuous
global symmetries are widely regarded as incompatible with string theory
and even the existence of gravity.
A different  possibility,
which does not seem to have been  mentioned before, might be
to suppose that the
PQ fields are actually moduli, making the origin a point of enhanced
symmetry. Then the non-renormalizable terms may be suppressed by a
factor $(\TeV/\mpl)^2 \sim 10^{-30}$. However this is small enough
for all $d\geq 5$ only if $f_{PQ}$ has an  implausibly small
value of order $10^8\GeV$. In any case, this mechanism cannot be used
in the flaton case, which invokes an unsuppressed coefficient for one
term.}

Because of these considerations, one may favour
a solution of the CP problem
in which the QCD axion is identified with a string axion \cite{bdaxion},
or one without any axion at all.\footnote
{The string axion as a curvaton candidate is discussed in \cite{pngb}.
As the string axion is a PNGB one may be hopeful that it can be kept
sufficiently light in the early Universe,  but string phenomenology is not
yet sufficiently developed that one can be sure.}
 In that case the field considered
in our proposal becomes ad hoc, introduced solely to explain the
curvature perturbation. Still, because of its  simplicity
and the ease with which it is made  sufficiently light,
we feel that the present curvaton model
is extremely attractive in comparison with
others. Let us end by mentioning those   other models
 which have a close connection with ours.

The closely  related  models are considered in
Refs.~\cite{kkt,McD}, where as in our case the effect of the
angular part of a flat direction has been considered. The
difference from our case is that the flat direction is supposed to
have zero VEV, so that the curvaton does not exist as a particle
in the vacuum. As a consequence, the phase can only induce the
curvature fluctuations in the radial field which leads to
different predictions.    In addition, the proposal of \cite{kkt}
does not invoke the mass-squared of order $-H^2$, so that the
radial potential is very flat with only a
 two-loop thermal correction breaking the
symmetry.
(In both cases, successful curvature
perturbations can arise for $n=3$, contrary to our case with
$n=1$.)  Because of such features the success of these models
depends on computations which are much more tricky than in our
case, though that is of course not necessarily an argument against
them. However, we note that
 the computation of
\cite{kkt}, involving the very flat radial potential,
 works only if inflation lasts for a limited amount of time,
because it takes the radial field during inflation to be at the
edge of the slow-roll regime, $V'' \sim H^2$. It seems more
reasonable to assume that inflation lasts long enough to allow the
quantum fluctuation to randomize the radial field  within  the
smaller  region  $V\lsim H^4$. (The analogous assumption for the
angular  field is of course the one that we made.)

 The other
related model \cite{pqcurv}
 also invokes a flaton model of PQ symmetry breaking,
but now the curvaton candidate is a radial field and it is not
clear how to keep it light in the early Universe.
 Also, the model works only if
 inflation is of short enough duration that the curvaton does not enter
the randomization regime.

\medskip

{\bf Acknowledgments:} EJC was supported by the Korea Research
Foundation Grant, KRF-2002-070-C00022.

\bigskip

\appendix

\section{Appendix: Randomization of the PNGB curvaton}

In this appendix we elaborate more on the assumption of
the complete randomization of the value of the curvaton during inflation.
Since the curvaton is an effectively massless
field during inflation, its value is perturbed by the action of quantum
fluctuations, which introduce a perturbation of the order of
\mbox{$\delta\sigma=T_H$} per Hubble time, where \mbox{$T_H=H_*/2\pi$}
is the Hawking temperature of de-Sitter space. Provided that the potential
$V(\sigma)$ is not too steep, these quantum
`kicks' can move the field around by random walk so that,
given enough e-foldings, the phase $\theta_0$ may assume an arbitrary value
(typically of order unity) by the time when the cosmological scales exit the
horizon. The criterion for this to occur for a PNGB curvaton is
\mbox{$H_*\geq H_c$}, where
\begin{equation}
H_c\equiv
\sqrt{m_\sigma v}.
\label{Hc}
\end{equation}

The above critical value has been obtained by considering the fact that the
region of $V(\sigma)$, where the field is randomized, is determined by the
condition \mbox{$V(\sigma)\lsim H_*^4$} \cite{prob}. This condition can be
understood as follows: The kinetic energy density of the quantum `kicks' is
\mbox{$T_H^4\sim H_*^4$}. Hence, the quantum fluctuations can displace the
field from its minimum only up to potential density \mbox{$V\sim T_H^4$}. For a
PNGB \mbox{$V(\sigma)\sim m_\sigma^2\sigma^2$}, which means that the borders of
the randomized region are at \mbox{$\sigma_Q\sim H_*^2/m_\sigma$}. If $H_*$ is
large enough then the randomized region includes the entire range of $\sigma$.
Setting \mbox{$\sigma_Q\rightarrow v$} we obtain the critical value of $H_*$,
shown in Eq.~(\ref{Hc}), over which the PNGB is fully randomized.

Using Eqs.~(\ref{vsare}) and (\ref{msare}) the bound \mbox{$H_*\geq H_c$}
translates into
\begin{equation}
C_A\leq\left(\frac{H_*}{M_P}\right)^{\frac{2n}{n+1}}\sim(3\pi\theta_0\zeta)^2,
\label{Cbound1}
\end{equation}
where we also used Eq.~(\ref{zeta}) with \mbox{$r\gsim 1$}, according to
Eq.~(\ref{results}). With \mbox{$\theta_0\sim{\cal O}(1)$} we see that complete
randomization of $\sigma_*$ requires
\mbox{$C_A<10^{-7}$}.
According to Eq.~(\ref{Cs}), this condition is satisfied provided that
$|I|\lsim 10^{-7}\mpl$.
\footnote{Note that, in this case, there is no problem with
\mbox{$C_A\geq A/H_*\sim 10^{-10}$}.}

If complete randomization is not realized then $\sigma_*$ may be much smaller
than $v$ during inflation. In this case, according to Eq.~(\ref{ris}), $r$ can
be smaller than unity and the results in Eq.~(\ref{results}) are modified.
In particular, the typical value of $\sigma_*$ is estimated as
\begin{equation}
\sigma_*\simeq\theta_0 v\sim\left\{
\begin{array}{ll}
v & \quad{\rm for}\quad H_*\geq H_c\\
H_*^2/m_\sigma & \quad{\rm for}\quad H_*<H_c
\end{array}
\right.\,,
\label{srandom}
\end{equation}
where we considered that, in the non-randomized case, the typical value is
\mbox{$\sigma_*\sim\sigma_Q$}. Inserting the above into Eq.~(\ref{zetais}) and
after some algebra it is easy to find
\begin{equation}
\tilde r\equiv\frac{r}{4+3r}=\pi\zeta\sqrt{\frac{v}{m_\sigma}}\times\left\{
\begin{array}{ll}
H_c/H_* & \quad{\rm for}\quad H_*\geq H_c\\
H_*/H_c & \quad{\rm for}\quad H_*<H_c
\end{array}
\right.\,.
\label{rtilde}
\end{equation}
It it easy to see that $\tilde r(r)$ is an increasing function of
$r$. Hence, the WMAP non-Gaussianity constraint reads:
\mbox{$\tilde r(r)>\tilde r(9\times 10^{-3})\sim 2\times
10^{-3}$}. For a successful PNGB curvaton, this constraint has to
be satisfied at least for the maximum allowed $\tilde r$.
Therefore, in view of Eq.~(\ref{rtilde}), we obtain
\begin{equation}
\tilde r_{\rm max}\simeq\pi\zeta\sqrt{v/m_\sigma}>2\times 10^{-3}.
\label{random}
\end{equation}
Using Eqs.~(\ref{vsare}), (\ref{msare}) and (\ref{zeta}), we get
\begin{equation}
\tilde r_{\rm max}^2\simeq\frac{\pi\zeta}{\sqrt{C_A}\,\theta_0}\;\tilde r
\quad\Rightarrow\quad\tilde r_{\rm max}<\frac{\pi\zeta}{\sqrt{C_A}\,\theta_0}.
\label{rtildemax}
\end{equation}
Combining Eqs.~(\ref{random}) and (\ref{rtildemax}) we obtain
\begin{equation}
\sqrt{C_A}<\frac{\pi\zeta}{(2\times 10^{-3})\,\theta_0}\simeq
0.08\;\theta_0^{-1},
\label{Cbound2}
\end{equation}
which is typically satisfied by a successful curvaton.


\input{epsf}

\begin{center}
\begin{figure}

\begin{picture}(10,10)
\put(20,-440){
\leavevmode
\hbox{%
\epsfxsize=6in
\epsffile{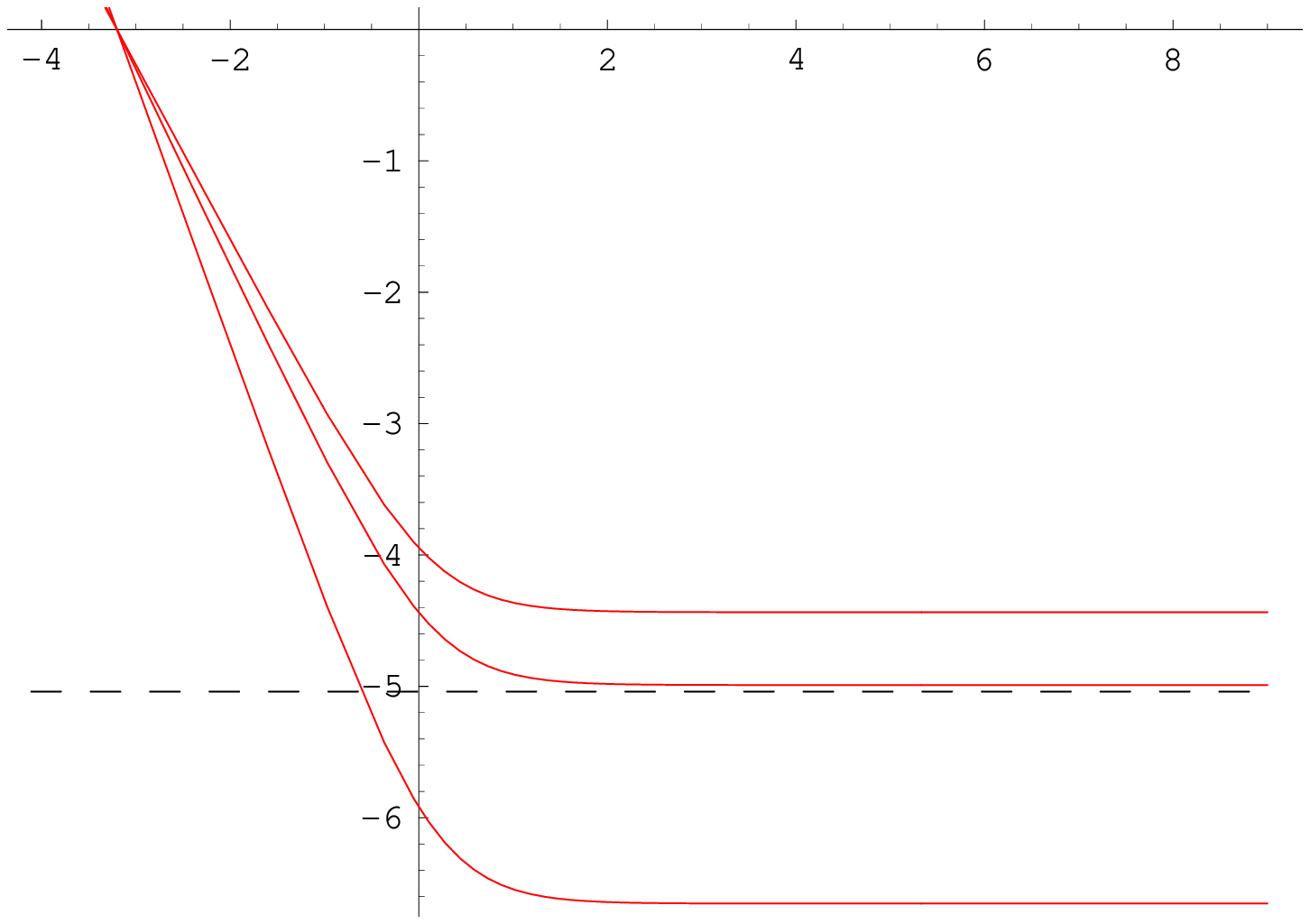}}
}
\put(350,30){\large $r$}
\put(110,-150){\large $H_*/M_P$}
\put(300,-81){\large $n=3$}
\put(300,-98){\large $n=2$}
\put(300,-148){\large $n=1$}
\end{picture}

\begin{picture}(10,10)
\put(22,-690){
\leavevmode
\hbox{%
\epsfxsize=6in
\epsffile{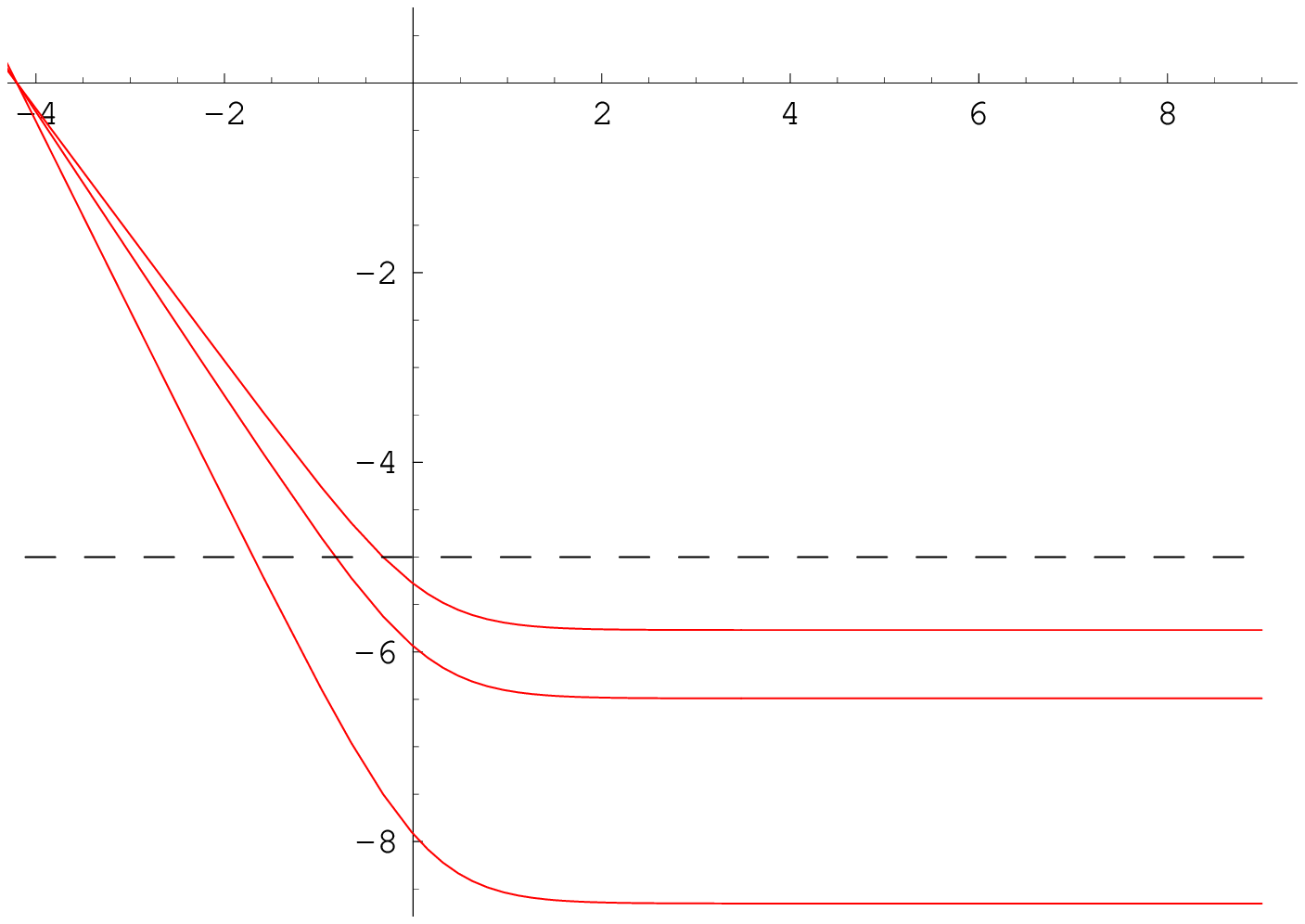}}
}
\put(345,-213){\large $r$}
\put(110,-378){\large $H_*/M_P$}
\put(300,-312){\large $n=3$}
\put(300,-328){\large $n=2$}
\put(300,-373){\large $n=1$}
\end{picture}

\vspace{15cm}
\caption{
Log-log plots of $H_*/M_P$ with respect to $r$ for all cases:
\mbox{$n=1,2,3$}. The dashed line corresponds to the bound:
\mbox{$H_*/M_P<10^{-5}$}. The upper graph corresponds to
\mbox{$\theta_0\sim 1$}, whereas the lower graph corresponds to
\mbox{$\theta_0\sim 0.1$}. We see that, in the former case, only
the case of \mbox{$n=1$} manages to avoid the bound. On the other
hand, in the latter case, all three possibilities are, in
principle, allowed. The curves meet when \mbox{$H_*\sim M_P$} as
expected.}

\end{figure}
\end{center}

\pagebreak

\begin{center}
\begin{figure}

\begin{picture}(10,10)
\put(20,-550){
\leavevmode
\hbox{%
\epsfxsize=5.5in
\epsffile{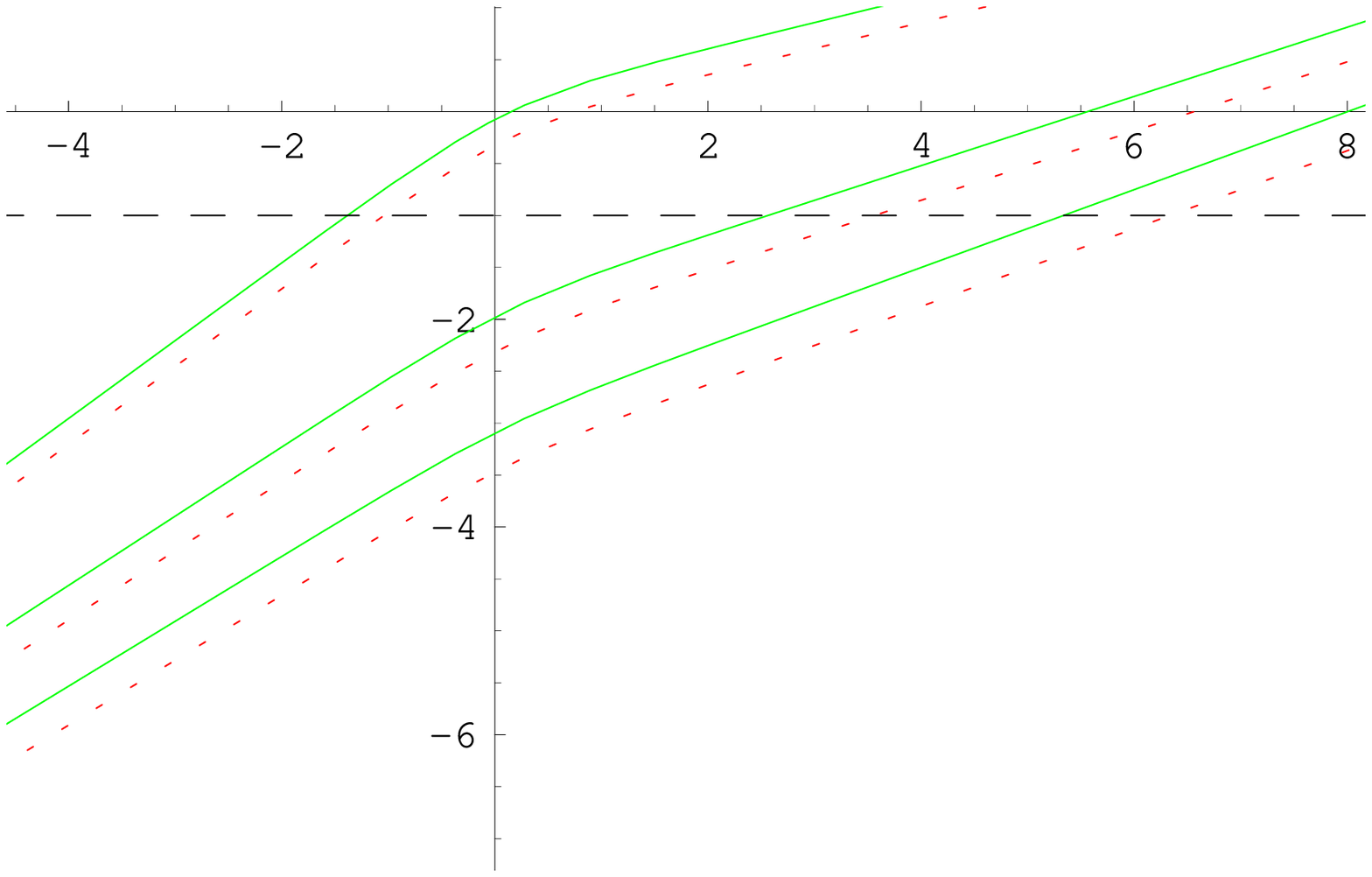}}
}
\put(70,-50){\large $r$}
\put(190,-240){\large $\theta_0$}
\put(400,-70){\large $n=3$}
\put(400,-50){\large $n=2$}
\put(290,-50){\large $n=1$}
\end{picture}

\vspace{10cm}
\caption{
Log-log plot of $\theta_0$ with respect to $r$ for all cases \mbox{$n=1,2,3$}.
The solid lines correspond to \mbox{$H_{\bar{m}}\sim 1$ GeV}, whereas the
dotted lines correspond to \mbox{$H_{\bar{m}}\sim m_{3/2}$}. The horizontal
dashed line corresponds to \mbox{$\theta_0\sim 0.1$}. We see that, for
\mbox{$H_{\bar{m}}>1$ GeV}, the value of $r$ is slightly larger, for a given
$\theta_0$ (it can increase up to an order of magnitude at most). Obviously, $\theta_0$ is bounded from above as
\mbox{$\log\theta_0<\log\pi\approx 0.5$}.
}
\end{figure}
\end{center}


\begin{center}
\begin{figure}
\vspace{-5cm}
\begin{picture}(10,10)
\put(-80,-1200){
\leavevmode
\hbox{\epsfxsize=13in
\epsffile{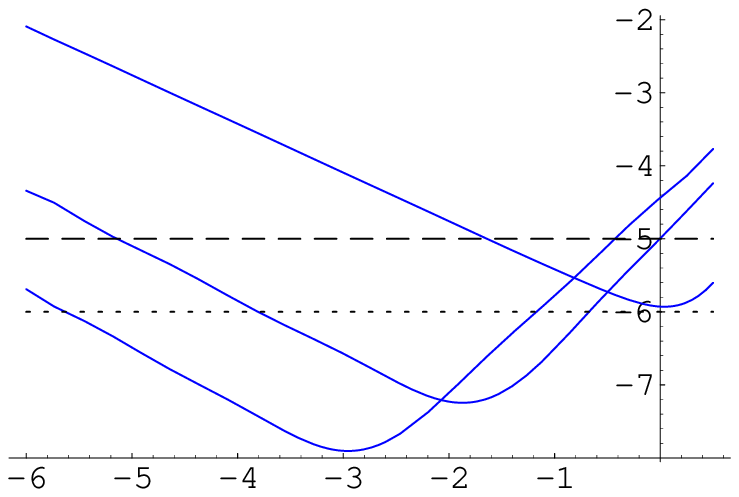}}
}
\end{picture}

\begin{picture}(10,10)

\put(380,-85){\Large $H_*/M_P$}
\put(85,-270){\Large $\theta_0$}
\put(85,-85){\Large $n=1$}
\put(85,-160){\Large $n=2$}
\put(85,-205){\Large $n=3$}

\end{picture}
\vspace{12cm}
\caption{
Log-log plots of $H_*/M_P$ with respect to $\theta_0$ for all cases:
\mbox{$n=1,2,3$} with \mbox{$H_{\bar{m}}\sim 1$ GeV}. The dashed line
corresponds to the bound: \mbox{$H_*/M_P<10^{-5}$} (which ensures the
compatibility with the observations), while the dotted line
corresponds to the bound: \mbox{$H_*/M_P<10^{-6}$} (which ensures that the
inflaton does not contribute significantly to the curvature perturbations).
We see that, when \mbox{$\theta_0\sim 1$}, only the case of \mbox{$n=1$} is
possible with
\mbox{$H_*/M_P
\stackrel{\mbox{$>$}}{\sim}
10^{-6}$},
which means that the inflaton
generated curvature perturbations are not negligible. For the curvaton to be
the sole source of the curvature perturbations one has to limit oneself to
the \mbox{$n=2,3$} cases, where \mbox{$\theta_0\ll 1$}.
}
\end{figure}
\end{center}

\end{document}